\documentclass[preprint,showpacs,preprintnumbers,amsmath,amssymb,groupaddress]{revtex4}
\usepackage{bbm}
\usepackage{amsfonts}

\usepackage{graphicx}
\usepackage{dcolumn}
\usepackage{bm}

\begin{document}


\title{Application of canonical Hamiltonian formulation to nonlinear light-envelope propagations
}

\author{Guo Liang}
\affiliation{Laboratory of Nanophotonic Functional Materials and Devices, South
China Normal University, Guangzhou 510631,China}
\author{Qi
Guo} \email{guoq@scnu.edu.cn} \affiliation{Laboratory of Nanophotonic Functional Materials and Devices, South China Normal University, Guangzhou
510631,China}

\date{\today}

\begin{abstract}
We first point out it is conditional to apply the variational approach to the nonlocal nonlinear Schr\"{o}dinger equation (NNLSE), that is, the response function must be an even function. Different from the variational approach, the canonical Hamiltonian formulation for the first-order differential system are used to deal with the problems of the nonlinear
light-envelope propagations. The Hamiltonian of the system modeled
by the NNLSE is obtained, which
can be expressed as the sum of the generalized kinetic energy and
the generalized potential. The solitons correspond to extreme points of the
generalized potential. The stabilities of solitons in both
local and nonlocal nonlinear media are also investigated by the
analysis of the generalized potential. They are stable when the
potential has minimum, and unstable otherwise.
\end{abstract}

\pacs{42.65.Tg; 42.65.Jx; 42.70.Nq}.

\maketitle
\section{Introduction}
The Hamiltonian viewpoint provides a framework for theoretical
extensions in many areas of
physics~\cite{Goldstein-book-05,menon-EJP,Maschke-IEEE,Escobar-Automatica,Buchdahl-book-93}.
In classical mechanics it forms the basis for further developments,
such as Hamilton-Jacobi theory, perturbation approaches and chaos. The canonical equations of Hamilton in classical mechanics are of the form
\begin{equation}\label{CE for 2ed}
\dot{q_i}=\frac{\partial H}{\partial p_i},
 -\dot{p_i}=\frac{\partial H}{\partial q_i},\quad (i=1,\cdots,n),
\end{equation}
where $q_i$ and $p_i$ are said to be the generalized coordinate and the generalized momentum, $\dot{q}_i=d q_i/dt$, $\dot{p}_i=d p_i/dt$, and $H$ is the Hamiltonian.
There are many situations in which the
Hamiltonian is equal to the sum of the generalized kinetic energy
$T$ and the generalized potential $V$.
The conditions are that the generalized potential is not a
function of generalized velocities, and the generalized
kinetic energy is a homogeneous quadratic function of
generalized velocities.
Once the generalized potential $V$, which is under the framework of
the Hamiltonian system, is obtained, the problem of the small
oscillations of a system about positions of equilibrium can be
easily dealt with. For a conservative mechanical system its
equilibrium state can be obtained by
$
\left(\frac{\partial V}{\partial q_i}\right)_0=0.
$
The generalized potential has an extremum at the
equilibrium configuration of the system, which is marked with the
subscript $0$. The equilibrium  is stable when the
extremum of the potential $V$ is the minimum, and unstable otherwise.

Up to now, to our knowledge, the canonical equations of Hamilton appearing in all the literatures are of the form (\ref{CE for 2ed}) except for our recent work, where it is pointed out that the canonical equations of Hamilton (\ref{CE for 2ed}) are only valid for the second-order differential
system (the system described by the second-order partial differential equation about the evolution coordinate) but not
valid for the first-order differential
system (the system described by the first-order partial differential equation about the evolution coordinate).
%
The nonlinear Schr\"{o}dinger equation (NLSE) is the first-order differential
system, which is a universal nonlinear
model that describes many nonlinear physical systems and can be
applied to hydrodynamics~\cite{Nore-Physica D-93}, nonlinear
optics~\cite{Anderson-pra-83}, nonlinear
acoustics~\cite{Bisyarin-AIP Conf. Proc.-08}, Bose-Einstein
condensates~\cite{seaman-pra-05}, and so on. The work about the canonical equations of Hamilton for the first-order differential
system
 will be introduced briefly here.

The
approximate analytical solutions of the NLSE can be obtained by the variational approach~\cite{Anderson-pra-83,Hasegawa-apl-73,Anderson-josab-1988,Malomed-progress in optics-2002},
where the light-envelope is treated as the classical
particle traveling in an equivalent potential, whose minimum corresponds to the soliton.
 In this paper we use the canonical
equations of Hamilton to deal with the nonlinear light-envelope
propagations. Such an approach is different from the the variational approach, which will be illustrated in the paper.
We can divide the Hamiltonian of the system
into the generalized kinetic energy and the generalized potential,
the extreme point of which corresponds to the soliton. But in some other
literatures~\cite{Seghete-pra-2007,Picozzi-prl-2011,Lashkin-pla-2007,Petroski-oc-2007},
solitons are regarded as the extrema of the Hamiltonian of
the system. Such a treatment has some problems, which will
be illustrated in the paper. To determine the stabilities of the soliton,
we can determine whether the generalized potential has a
minimum. Solitons are stable when the generalized potential
has a minimum, but unstable otherwise. In fact, the similar
expression, the Hamiltonian expressed as the sum of the generalized
kinetic energy and the generalized potential, appeared in
Ref.\cite{Desyatnikov-prl-10}, but the elaboration of the systematic
theoretical principle was absent, which is often of great
importance.

The paper is organized as follows. In Sec. \ref{for soliton}, we briefly introduce the model, the nonlocal nonlinear Schr\"{o}dinger
equation (NNLSE). The restriction on the the response function in the the variational approach is discussed in Sec. \ref{limition on response}, we point out the variational approach can be used to find the approximately analytical solution of the NNLSE if and only if the response function is an even function. We use the canonical
equations to deal with the nonlinear light-envelope
propagations in the paper, but the conventional canonical equations of Hamilton are not valid for NNLSE, so in Sec. \ref{ceh for 1st system} we will briefly introduce the canonical equations of Hamilton valid for the first-order differential
system, which will be reported elsewhere in detail. The application of the canonical equations of Hamilton introduced in Sec. \ref{ceh for 1st system} to the NNLSE is shown in Sec. \ref{application}. By use of the canonical equations of Hamilton, we can find the soliton solutions of the NNLSE, and can analyze the stability characteristics of the solitons. The difference between the variational approach and the approach employed in the paper is discussed in Sec. \ref{difference}. Sec. \ref{conclusion} gives the summary.

\section{Model}\label{for soliton}
 The propagation of the light-envelope in the nonlocal cubic
nonlinear media is modeled by the nonlocal nonlinear Schr\"{o}dinger
equation (NNLSE) in the dimensionless system
~\cite{Snyder-science-97,Mitchell-josab-99,Krolikowski-pre-01,Conti-prl-10}
\begin{equation}\label{NNLSE}
i\frac{\partial \varphi}{\partial
z}+\Delta_{\bot}\varphi+\varphi\int_{-\infty}^\infty
R(\textbf{r}-\textbf{r}')|\varphi(\textbf{r}',z)|^2d^D\textbf{r}'=0,
\end{equation}
where $\varphi(r,z)$ is the complex amplitude envelop, $z$ is the longitudinal coordinate, $\textbf{r}$ and
$\textbf{r}'$ are the $D$-dimensional transverse coordinate vectors.
$d^D\textbf{r}'$ is a $D$-dimensional volume element at
$\textbf{r}'$, $\Delta_{\bot}$ is the $D$-dimensional transverse
Laplacian operator, and $R$ is normalized response function of the media
such that $\int_{-\infty}^\infty R(\textbf{r}')d^D\textbf{r}'=1$.
For a singular response, i.e., $R(\textbf{r})=\delta(\textbf{r})$, Eq.(\ref{NNLSE}) simplifies
to the NLSE
\begin{equation}\label{NLSE}
i\frac{\partial \varphi}{\partial
z}+\Delta_{\bot}\varphi+|\varphi|^2\varphi=0.
\end{equation}
When $D=1$, NNLSE (\ref{NNLSE}) can describe the propagations of
both optical
beams~\cite{Snyder-science-97,Mitchell-josab-99,Krolikowski-pre-01}
and pulses~\cite{Conti-prl-10}. Particularly, it was predicted very
recently~\cite{Conti-prl-10} that strongly nonlocal temporal
solitons can exist in the model~(\ref{NNLSE}). The second term of
(\ref{NNLSE}) models the diffraction for the optical beam, and the
group velocity dispersion (GVD)~\cite{Agrawal-book-01} for the
optical pulse. The nonlinear term of (\ref{NNLSE}) describes the
self focusing of the optical beam~\cite{Chiao-prl-64} and optical
pulse~\cite{Hasegawa-apl-73}.
Generally speaking, when $D=2$, NNLSE (\ref{NNLSE}) only describes
the propagations of optical beams. The propagation of a pulsed
optical beam can be described by the NNLSE, and a optical
bullet~\cite{Silberberg-ol-90} can be obtained when $D=3$. For
$D>3$, the NNLSE (\ref{NNLSE}) is just a phenomenological model, the
counterpart of which can not be found in physics. The response function $R$ can
be symmetric for the optical beam, but is asymmetric for the optical pulse due to the causality~\cite{Boyd-book-92}.

\section{Discussion about the variational approach for the  nonlocal nonlinear Schr\"{o}dinger
equation}\label{limition on response}
To find the approximately analytical solution of the NNLSE, the variational approach is widely used~\cite{Guo-oc-06}. The reason that the variational approach can be applied to the NNLSE is that the NNLSE  can
be viewed as the Euler-Lagrange equation
\begin{equation}\label{euler-lagrange equation}
\frac{\partial}{\partial z}\frac{\partial l}{\partial\left(\frac{\partial\varphi^*}{\partial z}\right)}+\frac{\partial}{\partial x}\frac{\partial l}{\partial\left(\frac{\partial\varphi^*}{\partial x}\right)}-\frac{\partial l}{\partial\varphi^*}=0,
\end{equation}
where $l$ is the Lagrangian density. Replacing $\varphi^*$ with $\varphi$, the complex-conjugate equation of the NNLSE can be obtained from the Euler-Lagrange equation (\ref{euler-lagrange equation}). It is easy to calculate the first two terms of Eq.~(\ref{euler-lagrange equation}), but is some difficult to calculate the last term because of the convolution between the response function and the intensity of the optical beam for the NNLSE. In the following we will take the NNLSE (\ref{NNLSE}) with $D=1$ as an example to discuss the condition under which the NNLSE (\ref{NNLSE}) is equivalent to the Euler-Lagrange equation (\ref{euler-lagrange equation}).

The Lagrangian density of the NNLSE (\ref{NNLSE}) is \cite{Guo-oc-06}
\begin{equation}
l=\frac{i}{2}\left(\varphi^*\frac{\partial\varphi}{\partial
z}-\varphi\frac{\partial \varphi^*}{\partial
z}\right)-\left|\frac{\partial\varphi}{\partial x}\right|^2+\frac{1}{2}\left|\varphi(x,z)\right|^2\Delta
n,\label{Lagrangian density}
\end{equation}
where $\Delta n=\int_{-\infty}^\infty
R(\textbf{r}-\textbf{r}')\left|\varphi(\textbf{r}',z)\right|^2d^D\textbf{r}'$.
 Inserting the Lagrangian density (\ref{Lagrangian density}) into the Euler-Lagrange equation (\ref{euler-lagrange equation}), the first two terms of (\ref{euler-lagrange equation}) can be easily obtained as
 \begin{equation}\label{first two terms}
 \frac{\partial}{\partial x}\frac{\partial l}{\partial\left(\frac{\partial\varphi^*}{\partial x}\right)}+\frac{\partial}{\partial z}\frac{\partial l}{\partial\left(\frac{\partial\varphi^*}{\partial z}\right)}=i\frac{\partial \varphi}{\partial
z}+\frac{\partial^2\varphi}{\partial x^2}.
 \end{equation}
 To calculate the last term,
 we first construct a functional by integrating the last term of the Lagrangian density (\ref{Lagrangian density}) as
\begin{equation}
F(\varphi,\varphi^*)=\frac{1}{2}\int_{-\infty}^{\infty}\Delta n(x)|\varphi(x)|^2dx=\frac{1}{2}\int_{-\infty}^{\infty}\int_{-\infty}^{\infty}R(x-x')|\varphi(x')|^2|\varphi(x)|^2dx'.
\end{equation}
The variation of the functional $F(\varphi,\varphi^*)$ can be obtained by definition as
\begin{eqnarray}
\delta F(\varphi,\varphi^*)&=&\frac{\partial}{\partial \varepsilon}F(\varphi+\varepsilon\delta\varphi,\varphi^*+\varepsilon\delta\varphi^*)|_{\varepsilon\rightarrow0}\nonumber\\
&=&\frac{1}{2}\int_{-\infty}^{\infty}\int_{-\infty}^{\infty}R(x-x')|\varphi(x)|^2\left[\varphi(x')\delta\varphi^*(x')+\varphi^*(x')\delta\varphi(x')\right]dx'dx\nonumber\\
&&+\frac{1}{2}\int_{-\infty}^{\infty}\Delta n\left[\varphi(x)\delta\varphi^*(x)+\varphi^*(x)\delta\varphi(x)\right]dx.
\end{eqnarray}
If the response function is a even function, i.e., $R(x)=R(-x)$, then we can obtain that
$\int_{-\infty}^{\infty}\int_{-\infty}^{\infty}R(x-x')|\varphi(x)|^2\left[\varphi(x')\delta\varphi^*(x')+\varphi^*(x')\delta\varphi(x')\right]dx'dx=\int_{-\infty}^{\infty}\Delta n\left[\varphi(x)\delta\varphi^*(x)+\varphi^*(x)\delta\varphi(x)\right]dx$. Then the variation of the functional $F(\varphi,\varphi^*)$ is simplified to
\begin{equation}\label{deal with convolution}
\delta F(\varphi,\varphi^*)=\int_{-\infty}^{\infty}\Delta n\varphi(x)\delta\varphi^*(x)dx+\int_{-\infty}^{\infty}\Delta n\varphi^*(x)\delta\varphi(x)dx.
\end{equation}
Because the variation of the functional $F(\varphi,\varphi^*)$ can be also expressed as
\begin{equation}\label{variation}
\delta F(\varphi,\varphi^*)=\int_{-\infty}^{\infty}\frac{\partial}{\partial \varphi}\left[\frac{1}{2}\Delta n(x)|\varphi(x)|^2\right]\delta\varphi(x)dx+\int_{-\infty}^{\infty}\frac{\partial}{\partial \varphi^*}\left[\frac{1}{2}\Delta n(x)|\varphi(x)|^2\right]\delta\varphi^*(x)dx.
\end{equation}
Comparing Eq.(\ref{deal with convolution}) and (\ref{variation}), we obtain
\begin{eqnarray}
\frac{\partial}{\partial \varphi^*}\left[\frac{1}{2}\Delta n(x)|\varphi(x)|^2\right]&=&\Delta n\varphi(x),\label{tonnlse}\\
\frac{\partial}{\partial \varphi}\left[\frac{1}{2}\Delta n(x)|\varphi(x)|^2\right]&=&\Delta n\varphi^*(x).\label{tonnlsecm}
\end{eqnarray}
Then the NNLSE (\ref{NNLSE}) can be obtained from the  Euler-Lagrange equation (\ref{euler-lagrange equation}) by combining Eq.(\ref{first two terms}) and Eq.(\ref{tonnlse}), its complex-conjugate equation can be obtained by combining Eq.(\ref{first two terms}) and Eq.(\ref{tonnlsecm}).

Consequently, it is conditional to apply the variational approach to the NNLSE, that is, the response function must be an even function. When the response function is not an even function, the variational approach will do not work any longer.
\section{Canonical equations of Hamilton for the first-order differential system}\label{ceh for 1st system}
We will use the canonical
equations of Hamilton to deal with the nonlinear light-envelope
propagations. However the canonical equations of Hamilton appearing in all the literatures, except for our recent work, are only valid for the second-order differential
system but not
valid for the first-order differential
system while the NNLSE (\ref{NNLSE}) is the first-order differential
system. Therefore, it is necessary to briefly introduce the canonical equations of Hamilton for the first-order differential
system first.

For the first-order differential system of the continuous systems, the Lagrangian density must be the linear function of the generalized velocities, and expressed as
\begin{equation}\label{Lagrangian density for the fist-order system}
l=\sum_{s=1}^NR_s(q_s)\dot{q}_s+Q(q_s,q_{s,x}),
\end{equation}
where $R_s$ is not the function of a set of $q_{s,x}$. Consequently, the generalized momentum $p_s$, which is obtained by the definition $p_s=\partial l/\partial\dot{q}_s$ as
\begin{equation}\label{equations of p q}
p_s=R_s(q_s), (s=1,\cdots,N)
\end{equation}
is only a function of $q_s$. There are $2N$ variables, $q_s$ and $p_s$, in Eqs. (\ref{equations of p q}). The number of Eqs. (\ref{equations of p q}) is $N$, which also means there exist $N$ constraints between $q_s$ and $p_s$. So the degree of freedom of the system given by Eqs. (\ref{equations of p q}) is $N$. Without loss of generality, we take $q_1,\cdots,q_\nu$ and $p_1,\cdots,p_\mu$ as the independent variables, where $\nu+\mu=N$. The remaining generalized coordinates and generalized momenta can be expressed with these independent variables as
$
q_\alpha=q_\alpha(q_1,\cdots,q_\nu,p_1,\cdots,p_\mu)(\alpha=\nu+1,\cdots,N),$ and $
p_\beta=p_\beta(q_1,\cdots,q_\nu,p_1,\cdots,p_\mu) (\beta=\mu+1,\cdots,N).
$
The Hamiltonian density $h$ for the continuous system is obtained by the Legendre transformation
as
$
 h=\sum_{s=1}^N\dot{q}_sp_s-l
$, where the Hamiltonian density $h$ is a function of $\nu$ generalized coordinates, $q_1,\cdots,q_\nu$, and $\mu$ generalized momenta, $p_1,\cdots,p_\mu$. We can obtain $N$ canonical equations of Hamilton
{\setlength\arraycolsep{0pt}
\begin{eqnarray}
\frac{\delta h}{\delta q_\lambda}&=&\sum_{s=1}^N\left(\dot{q}_s\frac{\partial p_s}{\partial q_\lambda}-\dot{p}_s\frac{\partial q_s}{\partial q_\lambda}\right)+\sum_{\alpha=\nu+1}^{N}\frac{\partial}{\partial x}\frac{\partial h}{\partial q_{\alpha,x}}\frac{\partial f_\alpha}{\partial q_\lambda},\label{canonical equations 1 for 1st order }\\
\frac{\delta h}{\delta p_\eta}&=&\sum_{s=1}^N\left(\dot{q}_s\frac{\partial p_s}{\partial p_\eta}-\dot{p}_s\frac{\partial q_s}{\partial p_\eta}\right)+\sum_{\alpha=\nu+1}^{N}\frac{\partial}{\partial x}\frac{\partial h}{\partial q_{\alpha,x}}\frac{\partial f_\alpha}{\partial p_\eta}\label{canonical equations 2 for 1st order }
\end{eqnarray}
}($\lambda=1,\cdots,\nu$, $\eta=1,\cdots,\mu$, and $\nu+\mu=N$).
The canonical equations of Hamilton (\ref{canonical equations 1 for 1st order }) and (\ref{canonical equations 2 for 1st order }) can be easily extended to the discrete system, which can be expressed as
{\setlength\arraycolsep{0pt}
\begin{eqnarray}
\frac{\partial H}{\partial q_\lambda}&=&\sum_{s=1}^N\left(\dot{q}_s\frac{\partial p_s}{\partial q_\lambda}-\dot{p}_s\frac{\partial q_s}{\partial q_\lambda}\right),\label{canonical equations 1 for 1st order for discrete system }\\
\frac{\partial H}{\partial p_\eta}&=&\sum_{s=1}^N\left(\dot{q}_s\frac{\partial p_s}{\partial p_\eta}-\dot{p}_s\frac{\partial q_s}{\partial p_\eta}\right),\label{canonical equations 2 for 1st order for discrete system}
\end{eqnarray}
}
where $\lambda=1,\cdots,\nu$, $\eta=1,\cdots,\mu$, and $\nu+\mu=N$.

\section{Application in nonlinear light-envelope
propagations}\label{application}
Before the application of the canonical equations of Hamilton (\ref{canonical equations 1 for 1st order for discrete system }) and (\ref{canonical equations 2 for 1st order for discrete system}), we should firstly calculate the Hamiltonian by the Legendre transformation
reading
$
 H=\sum_{s=1}^N\dot{q}_sp_s-L
$,
where
the Lagrangian $L$
can be obtained as
$
L=\int_{-\infty}^\infty ld^D\textbf{r}.
$
The Lagrangian $L$ is a function of generalized coordinates, $\varphi,\varphi^*$
and generalized velocities, $\dot{\varphi},\dot{\varphi}^*$.
 It is clear that the Lagrangian is not
an explicit function of $z$, so the Hamiltonian of the system is
conservative. Now, we assume the light-envelop has a given
form, $\varphi=\varphi\left(q_1,\cdots,q_n\right)$, where
$q_1,\cdots,q_n$ are the parameters changing with $z$. It can be
regarded as the variables transformation, with which
 we
transform the coordinate system expressed by the set of generalized
coordinate $\varphi$ to the one expressed by another set of
generalized coordinates $q_1,\cdots,q_n$.

Here we assume the material response is
the Gaussian function
$
R(\textbf{r})=\frac{1}{(\sqrt{\pi}w_m)^D}\exp\left(-\frac{|\textbf{r}|^2}{w_m^2}\right),
$
 and the trial solution has the form,
$
\varphi(r,z)=q_A(z)\exp\left[-\frac{r^2}{q_w^2(z)}\right]\exp\left[iq_c(z)r^2+iq_{\theta}(z)\right],
$
where $q_A,q_{\theta}$ are the amplitude and phase of the complex
amplitude of the light-envelope, respectively, $q_w$ is the width of the light-envelope, $q_c$
is the phase-front curvature, and they all vary with
the propagation distance $z$. We obtain the Lagrangian
\begin{eqnarray}\label{integral Lagrangian}
L&=&-2^{-2-D}\pi^{D/2}q_A^2q_w^{-2+D}(w_m^2+q_w^2)^{-D/2}\nonumber\\
&&[-2q_A^2q_w^{2+D}
+2^{D/2}(w_m^2+q_w^2)^{D/2}(4D\nonumber\\
&&+4Dq_c^2q_w^4+Dq_w^4\dot{q}_c+4q_w^2\dot{q}_{\theta})],
\end{eqnarray}
which is a function of
generalized coordinates, $q_A,q_w,q_c$ and generalized velocities,
$\dot{q}_c,\dot{q}_{\theta}$, but not an explicit function of $z$.
The
generalized momenta can be obtained
\begin{eqnarray}
p_A&=&p_w=0,\label{p 0}\\
p_c&=&-2^{-2-\frac{D}{2}}D\pi^{D/2} q_A^2 q_w^{2+D},\label{qc momentum}\\
p_{\theta}&=&-\left(\frac{\pi}{2}\right)^{D/2}q_A^2
q_w^D.\label{p_theta}
\end{eqnarray}
The Hamiltonian of the system then can be
determined by Legendre transformation
\begin{eqnarray}\label{Hamiltonian}
H&=&2^{-1-D}\pi^{D/2}q_A^2q_w^{-2+D}(w_m^2+q_w^2)^{-D/2}[-q_A^2q_w^{2+D}\nonumber\\
&&+2^{1+\frac{D}{2}}D(w_m^2+q_w^2)^{D/2}(1+q_c^2q_w^4)],
\end{eqnarray} and can be proved to be a constant, i.e. $\dot{H}=0.$

There are four generalized coordinates and four generalized momenta in the four equations (\ref{p 0})(\ref{qc momentum})(\ref{p_theta}). So the degree of freedom of the set of equations (\ref{p 0})(\ref{qc momentum})(\ref{p_theta}) is four. Without loss of generality, we take $q_c,q_\theta,p_c$ and $p_\theta$ as the independent variables. From Eqs.(\ref{qc momentum})(\ref{p_theta}), the generalized
coordinates $q_A,q_w$ can be expressed by generalized momenta
$p_c$ and $p_\theta$ as
$
q_A=(-p_\theta)^{1/2}[Dp_\theta/(2\pi
p_c)]^{D/4},
q_w=[4p_c/(Dp_\theta)]^{1/2},
$
inserting which into the Hamiltonian
(\ref{Hamiltonian}), we have
\begin{eqnarray}\label{Hamiltonian with momenta}
H=-\frac{ D^2 p_{\theta }^2+16 p_c^2 q_c^2}{4 p_c}-\frac{1}{2} \pi
^{-D/2}(\frac{4 p_c}{D p_{\theta }}+w_m^2)^{-D/2}.
\end{eqnarray}
 By use of the canonical equations of Hamilton (\ref{canonical equations 1 for 1st order for discrete system }) and (\ref{canonical equations 2 for 1st order for discrete system}), where $\mu=\nu=2$ and $n=4$, we can obtain the following four
equations
\begin{eqnarray}
\dot{q}_c&=&\frac{D^2 p_{\theta }^2}{4 p_c^2}-4 q_c^2+\frac{D \pi
^{-D/2} p_{\theta }^2(\frac{4 p_c}{D p_{\theta }}+w_m^2)^{-D/2}}{4
p_c+D p_{\theta } w_m^2},\label{dqc}\\
\dot{q}_\theta&=&-\frac{(4+D) \pi ^{-D/2} p_c p_{\theta } (\frac{4 p_c}{D p_{\theta }}+w_m^2){}^{-D/2}}{4 p_c+D p_{\theta } w_m^2}\nonumber\\
&&-\frac{D^2p_{\theta }}{2p_c}-\frac{D \pi ^{-D/2} p_{\theta }^2
w_m^2 (\frac{4 p_c}{D p_{\theta }}+w_m^2){}^{-D/2}}{4 p_c+D
p_{\theta } w_m^2},\\
\dot{p}_c&=&8 p_cq_c,\label{dpc}\\
\dot{p}_\theta&=&0.\label{dp_theta}
\end{eqnarray}
%

Because $q_{\theta}$ is a cyclic coordinate, the
corresponding generalized momentum $p_{\theta}$ is a constant, which
can be confirmed by Eq.(\ref{dp_theta}). In fact, this also
represents that the power of the light-envelope,
$P_0=\int_{-\infty}^\infty\left|\varphi\right|^2d^D\textbf{r}=q_A^2(\sqrt{\pi/2}q_w)^D$,
is conservative. From this we can obtain
\begin{equation}\label{amplitude}
q_A^2=P_0(\sqrt{\pi/2}q_w)^{-D}.
\end{equation}
Taking the derivative with respect to $z$ on both sides of
Eq.(\ref{qc momentum}), then comparing it with Eq.(\ref{dpc}), we
can obtain with the aid of Eq.(\ref{amplitude})
\begin{equation}\label{qc}
q_c=\frac{\dot{q}_w}{4q_w}.
\end{equation}
Then inserting Eq.(\ref{qc}) into the Hamiltonian
(\ref{Hamiltonian}) with the aid of Eq.(\ref{amplitude}), we have
$
H=T+V,
$
where \begin{eqnarray} T&=&\frac{1}{16}DP_0\dot{q}_w^2,\\
V&=&\frac{D P_0}{q_w{}^2}-\frac{1}{2} \pi ^{-D/2} P_0^2
\left(w_m^2+q_w{}^2\right){}^{-D/2}\label{potential energy}.
\end{eqnarray}
For the system, there are only one independent generalized coordinate $q_w$ and one independent generalized velocity $\dot{q}_w$, which can be
proved in the Appendix. When the Hamiltonian is expressed with independent
variables, it is indeed the total energy expressed as
the sum of the generalized kinetic energy and the generalized potential.

The problem associated with the NNSLE is a problem of small oscillations from the
Hamiltonian point of view. The soliton corresponds to
the extremum point of the generalized potential. But in some
literatures~\cite{Seghete-pra-2007,Picozzi-prl-2011,Lashkin-pla-2007,Petroski-oc-2007},
solitons were regarded as the extrema of the Hamiltonian. Such a treatment has some problem, because in those
literatures the trial solution has a changeless
profile (solitonic profile), the system expressed with the solitonic profile is the static system. The kinetic energy of
the static system is zero, and the Hamiltonian is equal to
the potential of the static system. In this connection, the extremum of the
Hamiltonian is the extremum of the generalized potential of the static system. But when the
system deviates from the equilibrium, the extremum of the
Hamiltonian is not the extremum of the generalized potential.

From $\partial V/\partial q_w=0$, we have
\begin{equation}\label{critical power equation}
-\frac{32}{q_w^3}+8 \pi ^{-D/2} P_0
q_w\left(w_m^2+q_w^2\right)^{-1-\frac{D}{2}}=0.
\end{equation}
From Eq.(\ref{critical power equation}) we can easily obtain the
critical power, with which the light-envelope will propagate with a
changeless shape. Here we take the notation $P_c$ to denote the
critical power instead of $P_0$, then we obtain
\begin{equation}\label{critical power} P_c=\frac{4\pi^{D/2}\left(w_m^2+q_w^2\right)^{1+\frac{D}{2}}}{q_w^4}.
\end{equation}
When $P_0=P_c$, we can obtain that
$
\dot{q}_c=q_c=0\label{wavefront},$ which implies that the wavefront
of the soliton solution is a plane. The propagation constant is $
\dot{q}_\theta=[(4-D) q_w^2+4 w_m^2]/q_w^4
$.

 Then we elucidate the stability characteristics of the soliton
by means of the analysis of the generalized potential $V$.
Performing the second-order derivative of the generalized
potential $V$ with respect to $q_w$, then inserting the
critical power into it, we obtain
\begin{equation}\label{stability criterion}
\Upsilon\equiv\left.\frac{\partial^2V}{\partial q_w^2}\right|_{P_0=
P_c}=\frac{64 }{q_w^4}\left[2-\frac{2+D}{2
\left(1+\sigma^2\right)}\right],
\end{equation}
where $\sigma=w_m/q_w$ is the degree of nonlocality. The larger is
$\sigma$, the stronger is the degree of nonlocality. When
$\Upsilon>0$, the generalized potential has a minimum, and the soliton is
stable. From Eq.(\ref{stability criterion}) we can obtain the
criterion of the stability of solitons, that is
\begin{equation}\label{stability criterion2}
\sigma^2>\frac{1}{4}(D-2),
\end{equation}
which is, in fact, consistent with the Vakhitov-Kolokolov (VK)
criterion~\cite{Vakhitov-qe-75} with the aid of the results of
Ref.\cite{Bang-pre-02}, and is proved briefly in~\cite{prove}.
\subsection{The local case}
\begin{figure}[htb]
\centerline{\includegraphics[width=3.5cm]{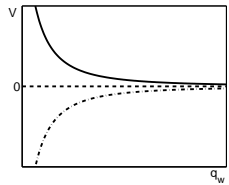}}
\caption{Qualitative plot of the potential $V$ as a function of
$q_w$ for three cases, $P_0<P_c$ (solid curve), $P_0=P_c$
(dashed curve), and  $P_0>P_c$ (dash-dot curve) when
$D=2$.}\label{potential plot}
\end{figure}

When $w_m\rightarrow0$, the response function
$R(\textbf{r})\rightarrow\delta(\textbf{r})$. The NNLSE will be reduced to the
NLSE (\ref{NLSE}).
Eqs. (\ref{critical power}) and (\ref{stability criterion}) are transformed to
\begin{equation}
P_c=4\pi^{D/2}q_w^{D-2},
\Upsilon=\frac{32}{q_w^4}(2-D).
\end{equation}
When $D=1$, $P_c=4\sqrt{\pi}/q_w$,
which is consistent with Eq.(42) of Ref.~\cite{Anderson-pra-83}. When $D=2$, $P_c=4\pi$, which is
the same as Eq.(16a) of Ref.~\cite{Desaix-josab-91}. We can
obtain $\Upsilon>0$ when $D<2$, $\Upsilon<0$ when $D>2$, and
$\Upsilon=0$ when $D=2$. So for the local case, the soliton is
stable for (1+1)-dimensional case, but  unstable when $D>2$. It
needs the further analysis for $D=2$ because $\Upsilon=0.$ When
$D=2$, the potential~(\ref{potential energy}) is deduced to
\begin{equation}\label{local potential energy}
V=\frac{\left(4\pi-P_0\right)P_0}{2\pi q_w^2},
\end{equation}
which
has no extreme when $P_0\neq 4\pi$. When $P_0=p_c=4\pi$, $V=0$, which is the extreme but not the minimum. So the
(1+2)-dimensional local solitons are unstable. The
relation between the potential $V$ and the width $q_w$ of the light-envelope is shown
in Fig.\ref{potential plot}. If the power of the light-envelope equals to
the critical power, the potential will be a constant, as can be seen
by dash curve of Fig.(\ref{potential plot}). Without the external
disturbance, the light-envelope will stay in its initial state, and keep its
width changeless. If the external disturbance makes the power
larger than the critical power, then the width will
become more and more smaller, and collapses at last, as can be
confirmed by the dash-dot curve of Fig.\ref{potential plot}. If the
external disturbance makes the power smaller than the
critical power, then the width will become more and more
larger, and diffracts at last, as can be confirmed by the solid
curve of Fig.\ref{potential plot}. These conclusions are consist
with those of Refs.~\cite{Berge-PR-98,Moll-prl-03,sun-oe-08}.
\subsection{The nonlocal case}
For the nonlocal case, when $D\leq2$, the condition (\ref{stability
criterion2}) can be satisfied automatically. That is to say the
(1+1)-dimensional and the (1+2)-dimensional nonlocal solitons are
always stable when the response function of the material is a
Gaussian function. It is consistent with the conclusion of
Ref.~\cite{Bang-pre-02}.
When $D>2$ the solitons can be stable only if the criterion of the
stability Eq.(\ref{stability criterion2}) should be satisfied first,
which is also the same as the result of Ref.\cite{Bang-pre-02}.

\section{differences from the variational
approach}\label{difference} The method employed in the paper, based
on the Hamiltonian formulation, is different from the widely used
variational approach. The first difference is that the equations
obtained by the two approaches are different. The equations obtained
by the variational approach are the differential
equations~\cite{Anderson-pra-83}. But the equation
we obtained is just a simple algebraic equation by differentiating
the generalized potential with respect to the generalized
coordinates. The other difference is that the "potentials" obtained
by the two approaches are very different. The potential obtained by
the variational approach~\cite{Anderson-pra-83} is just an
equivalent potential, which is obtained by comparing the evolution
of the width of the light-envelope with the motion of a particle in a
potential well. But in the paper, the potential we obtained is the
potential from the Hamiltonian point of view.

\section{Conclusion}\label{conclusion}
We point out the variational approach can be used to find the approximately analytical solution of the NNLSE if and only if the response function is an even function. We apply the canonical Hamiltonian formulation to nonlinear light-envelope propagations. The Hamiltonian of the nonlinear system can be expressed as the sum of the generalized
kinetic energy and the generalized potential.
Solitons correspond to the extreme of the generalized
potential. Solitons are stable when the generalized
potential has the minimum, and unstable otherwise.
\section*{ACKNOWLEDGMENTS}
This research was supported by the National Natural Science
 Foundation of China (Grant Nos. 11074080 and 10904041), the Specialized Research Fund for the Doctoral Program
 of Higher Education (Grant No. 20094407110008), and the Natural Science Foundation
of Guangdong Province of China (Grant No. 10151063101000017).

\section*{APPENDIX. Proof of a proposition}
If the Lagrangian of a system is expressed as
$L(q_1,\cdots,q_{j-1},q_{j+s},\cdots,q_n,\dot{q}_1,\cdots,\dot{q}_{i-1},\dot{q}_{i+m},\cdots,\dot{q}_n)$,
where $m$ generalized velocities,
$\dot{q}_i,\dot{q}_{i+1},\cdots,\dot{q}_{i+m-1}$, and $s$
generalized coordinates, $q_j,q_{j+1},\cdots,q_{j+s-1}$, are both
not included, then the system only has $n-m-s$ independent
variables($n-m-s$ independent generalized coordinates and $n-m-s$
independent generalized velocities).

Inserting the cyclic coordinates, $q_j,q_{j+1},\cdots,q_{j+s-1}$,
into the Euler-Lagrange equations (\ref{EulerLag}) and replacing $t$
with $z$, we have
$
\frac{d}{d z}\left(\frac{\partial L}{\partial
\dot{q}_\nu}\right)=0,\ (\alpha=1,\cdots,s),
$
i.e.
\begin{equation}\label{appendix cyclic coordinate equation}
\frac{\partial L}{\partial \dot{q}_\alpha}=C_\alpha,
\end{equation}
where $C_\alpha$ is a constant independent of $z$. Because NNLSE is a
first-order differential equation, the Lagrangian of the system
(\ref{integral Lagrangian}) is a function of the first degree in
$\dot{q}_\nu$, then from Eq.(\ref{appendix cyclic coordinate
equation}), we have the following $s$ constraints
\begin{equation}\label{appendix f}
f_\alpha(q_1,\cdots,q_{j-1},q_{j+s},\cdots,q_n)=0,
\end{equation}
where $f_\alpha=\partial L/\partial \dot{q_\alpha}-C_\alpha$. Form the
Euler-Lagrange equations associated with the disappearing
generalized velocities,
$\dot{q}_i,\dot{q}_{i+1},\cdots,\dot{q}_{i+m-1}$, we obtain another
$m$ constraints
\begin{equation}\label{appendix g}
g_\beta(q_1,\cdots,q_{j-1},q_{j+s},\cdots,q_n,\dot{q}_1,\cdots,\dot{q}_{i-1},\dot{q}_{i+m},\cdots,\dot{q}_n)\nonumber\\
=0,
\end{equation}
where $g_\beta=\partial L/\partial q_\beta$ and $\beta=1,\cdots,m.$ The remaining generalized
coordinates and generalized velocities of the Lagrangian appear in
pairs. They satisfy the differential equations
\begin{equation}\label{appendix j}
J_\gamma(q_1,\cdots,q_{j-1},q_{j+s},\cdots,q_n,\dot{q}_1,\cdots,\dot{q}_{i-1},\dot{q}_{i+m},\cdots,\dot{q}_n)=0,
\end{equation}
where $J_\gamma=\frac{d}{d z}(\frac{\partial
L}{\partial\dot{q}_\gamma})-\frac{\partial L}{\partial q_\gamma}$ and $\gamma=1,\cdots,n-m-s$.
Taking the derivative with respect to $z$ on both sides of
Eq.(\ref{appendix f}), we have
$$
F_\alpha(q_1,\cdots,q_{j-1},q_{j+s},\cdots,q_n,\dot{q}_1,\cdots,\dot{q}_{j-1},\dot{q}_{j+s},\cdots,\dot{q}_n)=0,
$$
where $F_\alpha=\sum_\iota\frac{\partial f_\alpha}{\partial q_\iota}\dot{q}_\iota$, and $\iota=1,\cdots,j-1,j+s,\cdots,n.$
Any $s$ generalized velocities in the function $F_\alpha$ can be
expressed with the remaining generalized velocities and all
generalized coordinates appearing in $F_\alpha$. Inserting the $s$
generalized velocities into the differential equations
(\ref{appendix j}), then there are only $n-m-s$ independent
generalized velocities appearing in (\ref{appendix j}). In a similar
way, any $m$ generalized coordinates in the function $g_\nu$ can be
expressed with the remaining generalized coordinates and all
generalized velocities appearing in $g_\nu$. Inserting the $m$
generalized coordinates into the differential equations
(\ref{appendix j}), then there are only $n-m-s$ independent
generalized coordinates appearing in (\ref{appendix j}).
Accordingly, there are only $n-m-s$ generalized coordinates and
$n-m-s$ generalized velocities for the system.


\begin{thebibliography}{99}
\bibitem{prove} To avoid confusion,
we make the notations used in Ref.\cite{Bang-pre-02} consistent with
those in the paper, where we use $w_m$ and $q_w$ to represent the
width of the response function and the light-envelope instead of
$\sigma,\beta$, respectively. In Ref.\cite{Bang-pre-02},
$P_s=(\pi/2)^{D/2}\alpha^2q_w^D$, with
$\alpha^2=(\lambda+D/q_w^2)(2+2w_m^2/q_w^2)^{D/2}$ and
$q_w^2=\left[4-D+\sqrt{(4-D)^2+16\lambda w_m^2}\right]/(2\lambda)$,
where $P_s$ is the soliton power, and $\lambda$ is the propagation
constant of solitons. According to the VK criterion, when
$dP_s/d\lambda>0$ the soliton become linearly stable.
We can derive
$\sigma^2>\frac{1}{4}(D-2)$ from the VK criterion $dP_s/d\lambda>0$,
where $\sigma=w_m/q_w$ represents the degree of nonlocality.
\end{thebibliography}
\end{document}